\documentclass[journal=nalefd,manuscript=letter]{achemso}

\usepackage[version=3]{mhchem} 
\usepackage{epstopdf}

\author{Enrique Burzur\'{\i}}
\email{e.burzurilinares@tudelft.nl}
\phone{+31 15 278 6099}
\affiliation{Kavli Institute of Nanoscience, Delft University of Technology, P. O. Box 5046, 2600 GA Delft, The Netherlands}

\author{Yoh Yamamoto}
\affiliation{Department of Physics, Virginia Tech, Blacksburg, Virginia 24061, USA}
\author{Michael Warnock}
\affiliation{Department of Physics, Virginia Tech, Blacksburg, Virginia 24061, USA}
\author{Xiaoliang Zhong}
\affiliation{Department of Physics, Virginia Tech, Blacksburg, Virginia 24061, USA}
\author{Kyungwha Park}
\affiliation{Department of Physics, Virginia Tech, Blacksburg, Virginia 24061, USA}
\alsoaffiliation{Kavli Institute of Nanoscience, Delft University of Technology, P. O. Box 5046, 2600 GA Delft, The Netherlands}
\author{Andrea Cornia}
\affiliation{Department of Chemical and Geological Sciences and INSTM, University of Modena and Reggio Emilia, via G. Campi 183, I-41125 Modena, Italy}
\author{Herre S. J. van der Zant}
\affiliation{Kavli Institute of Nanoscience, Delft University of Technology, P. O. Box 5046, 2600 GA Delft, The Netherlands}

\title{Franck-Condon Blockade in a Single-Molecule Transistor}

\abbreviations{IR,NMR,UV}
\keywords{Single-molecule magnets, molecular spintronics, Franck-Condon blockade, electron-vibron coupling}

\begin{document}




\begin{abstract}
We investigate vibron-assisted electron transport in single-molecule transistors containing an individual Fe$_{4}$ Single-Molecule Magnet. We observe a strong suppression of the tunneling current at low bias in combination with vibron-assisted excitations. The observed features are explained by a strong electron-vibron coupling in the framework of the Franck-Condon model supported by density-functional theory.
\end{abstract}


Vibrational modes (vibrons) play an essential role in the mechanics of a wide variety of nanostructures. In addition, they can couple to the electric charge, affecting the electrical transport through such nanoelectromechanical (NEMS) systems. Electron-vibron coupling has been experimentally observed, for instance, as vibron-assisted transport excitations in carbon nanotubes\cite{LeRoy2004,Sapmaz2006,Leturcq2009,Huttel2009}, in single molecules embedded in a solid-state transistor\cite{Park2000,Park2002,Pasupathy2005,Leon2008,Osorio2010} or probed by a scanning tunneling microscope (STM) configuration\cite{Stipe1998,Pradhan2005}. When vibrational modes are mechanically excited in NEMS, they may induce mechanical instabilities\cite{Pistolesi2007, Weick2010, Weick2011}. In magnetic molecules, the molecular vibrations may couple to spin degrees of freedom and play an important role in the molecular spin relaxation\cite{Gatteschi2006}. Very recently, experimental evidence of such spin-vibron coupling was reported for a single-molecule magnet (SMM) TbPc$_{2}$ grafted onto a carbon nanotube\cite{Ganzhorn2013}. In that case, the reversal of the SMM magnetic moment via an external magnetic field was indirectly observed in the conductance map from the coupling to vibrational excitations of the nanotube. Therefore, electron-vibron coupling may be in principle used to detect the magnetic states of nanostructures and, conversely, to manipulate their transport and magnetic properties.

An interesting feature induced by electron-vibron coupling is the Franck-Condon (FC) blockade effect which occurs when electric charge is strongly coupled to vibrations. A manifestation of the FC blockade effect is that single-electron tunneling is suppressed at low bias for any gate voltage\cite{Koch2005,Koch2006}. It was first observed by Weig et al.\cite{Weig2004} in electron transport through GaAs/AlGaAs quantum dots of several hundreds of nm. The FC blockade effect was later observed in suspended carbon nanotubes\cite{Sapmaz2006,Leturcq2009} that are typically orders of magnitude larger than individual molecules. So far, however, the FC blockade effect has not been systematically analyzed in single-molecule junctions despite several experimental studies focusing on electron-vibron coupling.

In this letter, we show experimental evidence of the FC blockade effect in electron transport via an individual magnetic molecule and present supporting calculations from density-functional theory (DFT). We investigate sequential electron tunneling (SET) through a SMM Fe$_{4}$ \cite{Accorsi2006} in a three-terminal configuration shown in Figure 1. We observe a dramatic suppression of current at low bias in combination with evenly spaced lines parallel to the Coulomb diamond edges in the conducting region. The suppression of current cannot be lifted by a gate voltage. The energy spacing of the excitations is 2.6~meV and is not affected by an applied magnetic field, thereby ruling out a possible magnetic origin. The energy spacing and the estimated resultant electron-vibron coupling constant are consistent with our DFT-calculated values. In addition, the DFT calculations suggest that the electron-vibron coupling can be enhanced by avoiding chemical bonding at the interface. Our findings clearly indicate strong electron-vibron coupling in magnetic molecules at the single-molecule level. They also open a more general route to investigate new rich physics in single-molecule transistors containing individual magnetic molecules; examples include studies of the effects of intrinsic molecular vibrations on quantum interferences\cite{Zhong2009} and on magnetic relaxation as a source of decoherence for SMM-based qubits\cite{Stamp2009}.


The structure of the Fe$_4$ SMM\cite{Accorsi2006} with formula [Fe$_4$(L)$_2$(dpm)$_6$] is shown in Figure 1a. The Fe$_{4}$ molecule consists of four Fe$^{3+}$ ions encapsulated in a hydrophobic shell made up of tert-butyl and phenyl groups from dpm$^{-}$ and L3$^{-}$ ligands, respectively. One Fe$^{3+}$ ion at the center is antiferromagnetically coupled with three Fe$^{3+}$ ions at the vertices of a triangle via oxygen bridges. The total spin in the ground state is $S=5$ and a magnetic anisotropy barrier of 16~K must be overcome to reverse the magnetic moment. The size of the Fe$_{4}$ SMM is 1.90 nm along the direction defined by the phenyl rings.

\begin{figure}
\includegraphics[width=1\textwidth]{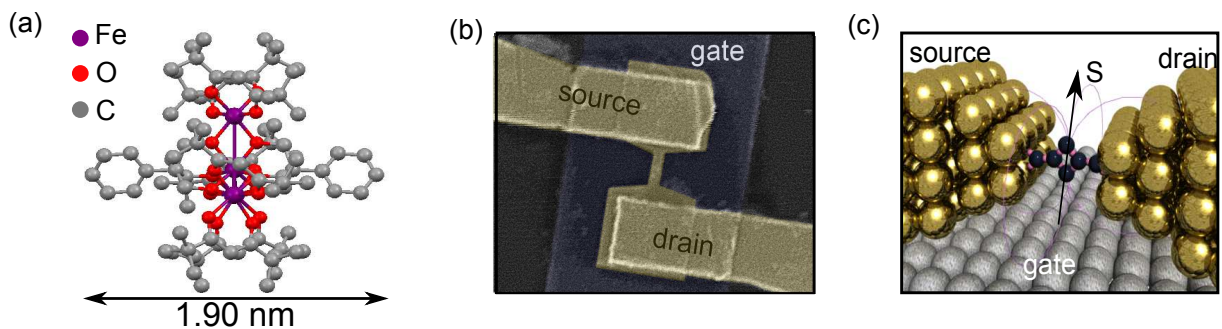}
\caption{\textbf{Fe$_{4}$ Single-molecule transistor.} (a) Sketch of the Fe$_{4}$ SMM. The magnetic core is made of 4 Fe$^{3+}$ ions (purple) surrounded by an organic shell (grey and red). Hydrogen atoms are not shown for clarity. The size of the molecule from ring to ring is 1.90 nm. (b) SEM image and (c) sketch of a molecular three-terminal transistor. An individual molecule is linked to gold source and drain electrodes that electrically bias the molecule. An underlying gate electrode is used to tune the levels of the molecule independently from the bias voltage (gate oxide separating gate from source/drain electrodes not shown).}
\label{figure1}
\end{figure}

A representative scanning electron microscope (SEM) image of the three-terminal device configuration is shown
in Figure 1b. Electromigration is used to thin the Au wire and followed by self-breaking in a solution of Fe$_{4}$ molecules to complete the device. Figure 1c schematically shows the sample layout: a single Fe$_4$ molecule is electrically linked to source and drain Au electrodes in order to apply a bias voltage to it. The Fe$_{4}$ molecule is not functionalized with specific surface-binding groups so that van der Waals interactions are responsible for the molecule-electrode coupling. An underlying gate electrode is used to tune the levels of the molecule independently from the bias voltage, as illustrated in Figure 1c. Importantly, our previous measurements showed that the magnetic structure of Fe$_{4}$ is preserved in a three-terminal configuration\cite{Burzuri2012}. Measurements are performed at 1.8~K unless specified otherwise.

Figure 2a shows a differential conductance color map in which $dI/dV$ is plotted as a function of bias $V$ and gate voltage $V_{g}$. Low-conductance regions (left and right blue areas) are indicative of two different charge states $N$ and $N+1$ which are accessible using the gate voltage. In this Coulomb blockade regime, the charge is stabilized within the molecule. Note that only two charge states are available, indicative of high addition energies as expected for a SMM. Strong high-conductance resonances, indicating SET through the molecule, separate adjacent charge states. The coupling of the molecule to the electrodes ($\Gamma_{L}$, $\Gamma_{R}$) is of the order of 1~meV and is obtained from the full width at half maximum (FWHM) of the Coulomb edges.

Interestingly, two remarkable features are observed in the $dI/dV$ map. First, SET is highly suppressed at low bias below a threshold value $V_{th}=\pm7.4$ meV. As a result, the Coulomb diamond edges do not intersect at zero bias, that is, no $dI/dV$ peak is observed at zero bias. This low-bias gap cannot be lifted by sweeping the gate voltage. Second, evenly spaced lines parallel to the Coulomb diamond edges are observed at positive and negative bias for $|V| > |V_{th}|$ (see Figure 2a). The lines become more visible when a numerical derivative of the $dI/dV$ is taken as shown in Figure 2b. The energy spacing between these excitations is $\Delta E=2.6$ meV. In the presence of a magnetic field, the Coulomb diamonds shift in gate as a result of the magnetic properties of the Fe$_{4}$ SMM. Interestingly, the value of $\Delta E$ is independent of applied magnetic field and therefore we rule out a magnetic origin for the excitations (see Supp. Info. for details). Moreover, the value of $\Delta E$ is symmetric with respect to the bias polarity and it is independent of the charge state, which is fingerprint of a vibronic origin. We emphasize that these two features with similar values of $\Delta E$ are observed for two additional junctions, each of which contains a single Fe$_{4}$ molecule. A discussion on the features of these additional junctions are presented in the Supporting Information. However, not all the Fe$_{4}$ molecular junctions show FC blockade as we discuss later.

\begin{figure}
\includegraphics[width=1\textwidth]{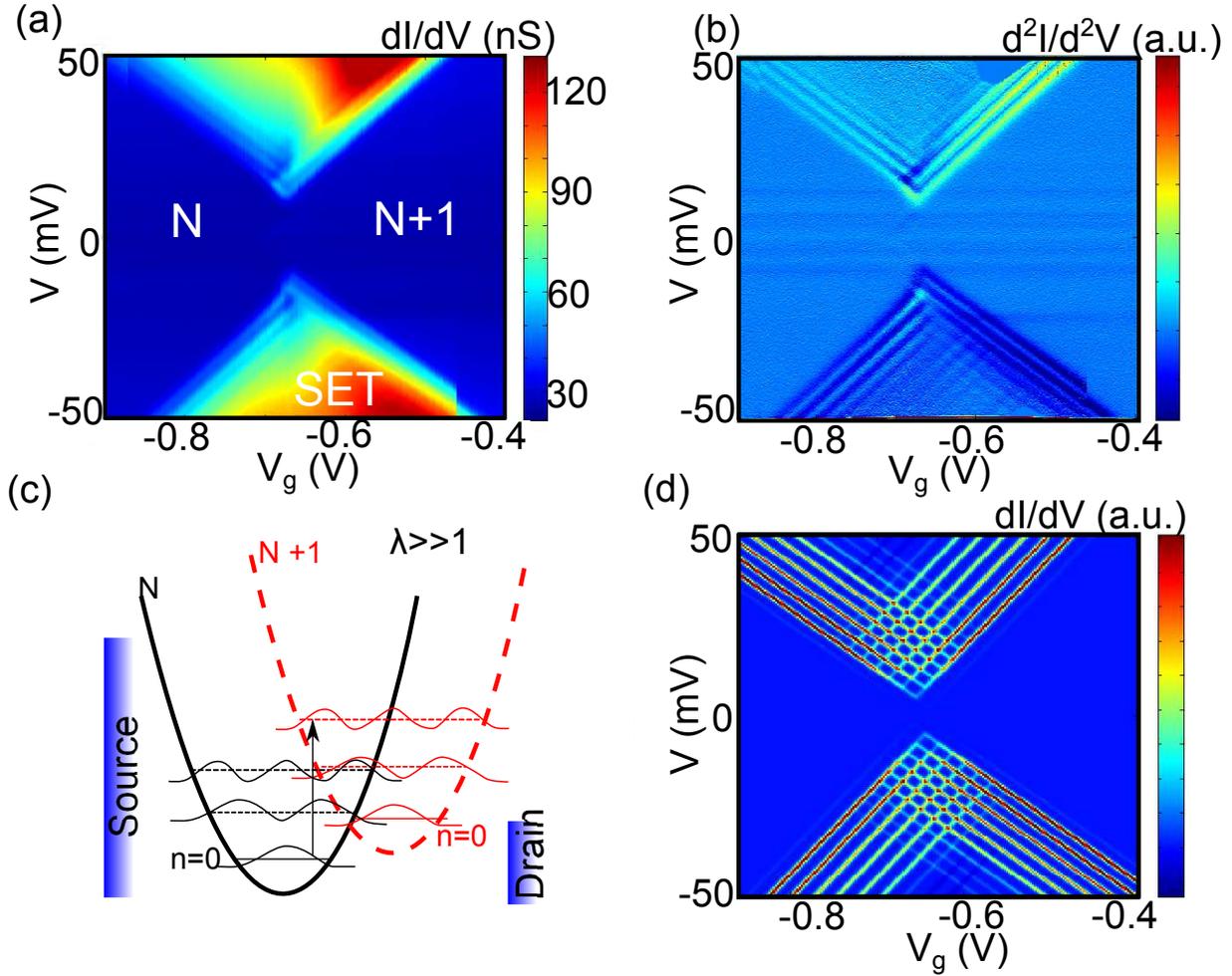}
\caption{\textbf{Franck-Condon blockade in a molecular junction containing a single Fe$_{4}$ molecule.} (a) Differential conductance ($dI/dV$) color map measured in an Fe$_{4}$ molecular junction as a function of bias $V$ and gate voltage $V_{g}$. Low-bias sequential-electron tunneling (SET) is suppressed below $V_{th}=7.4$ meV and the Coulomb blockade cannot be lifted by $V_{g}$. Periodic excitations appear within the SET regime at positive and negative bias. The energy spacing between excitations is $\Delta E=2.6$ meV. (b) Numerical derivative of the $dI/dV$ color plot shown in (a). Periodic excitations running parallel to the Coulomb diamond edges become more visible. (c) Schematic representation of the Franck-Condon model for strong electron-vibron coupling $\lambda$. For high values of $\lambda$, the equilibrium coordinates in adjacent charge states are significantly shifted from each other. Then, vibronic ground state to ground state transitions become exponentially suppressed but ground-state to excited-states become available when $V$ matches the energy of the excited vibrons ($n\hbar\omega$). (d) Calculated $dI/dV$ color map by introducing the electron-vibron coupling in the rate equations\cite{Seldenthuis2008}. The values of the temperature and the vibron energy used in the model are those obtained from the experiment ($T=1.8$ K and $\Delta E\sim\hbar\omega_{0}=2.6$ meV). The best agreement is obtained for $\lambda=2.21$.}
\label{figure2}
\end{figure}

The zero-bias conductance suppression may originate from the FC blockade effect\cite{Koch2005,Koch2006B} which occurs when the dimensionless electron-vibron coupling $\lambda$ is strong, i.e. $\lambda \gg 1$. So far, direct experimental evidence of the effect has been reported for carbon nanotubes \cite{Sapmaz2006,Leturcq2009} and semiconductor quantum dots\cite{Weig2004}, which are about 100 times larger in size than Fe$_{4}$. Figure 2c illustrates the FC blockade model. For a system with a large value of $\lambda$, the equilibrium coordinates of the electronic ground state greatly differ from those of an electronic excited state. In the present case of Fe$_{4}$, this corresponds to the case that the equilibrium geometry of the $N$ charge state is very different from that of the $N+1$ charge state, as sketched in Figure 2c. A transition from the vibrational ground level $n=0$ of the $N$ state to the vibrational ground level of the $N+1$ state is exponentially suppressed with $\lambda$\cite{Koch2005}. The low-bias gap observed in Figure 2b is due to the suppression of this transition. However, the probability for transitions to occur from the $n=0$ level of the $N$ state to the $n \neq 0$ levels of the $N+1$ state increases with a FC factor\cite{Koch2006B}
\begin{eqnarray}
{\cal F}_{n,0} &=& \frac{\lambda^{2n}}{n!} e^{-\lambda^{2}},
\label{eq:FC}
\end{eqnarray}
where $n$ is a vibrational quantum number with frequency $\omega_{0}$, assuming that only one vibrational mode is considered. Note that the FC factor only depends on the dimensionless quantities $\lambda$ and $n$.

Transitions involving higher-energy vibrational levels ($n\neq0$) start to contribute to the sequential tunneling when $n > \lambda^2/(2 \log \lambda) \sim \lambda^2$ (see also Eq.~(\ref{eq:FC})). The observed parallel lines in the SET region in Figure 2b are due to such transitions and therefore the energy spacing $\Delta E$ of the lines corresponds to $\hbar \omega_{0}$. The suppression of the conductance is predicted to be prominent for equilibrated vibrons with zero relaxation time \cite{Koch2005} and for $k_{B}T \ll \hbar\omega_{0}$, which meets our experimental conditions. The FC blockade is lifted when the bias voltage matches a multiple of vibrational energy, $n \hbar \omega_0$, at a threshold bias voltage $V_{th}$ of about $\lambda^2 \hbar\omega_{0}$ \cite{Koch2005}. Using the experimental values of $\hbar\omega_0$ and $V_{th}$, we extract the value $\lambda$=1.7.

To corroborate our interpretation of the observed conductance map we simulate a $dI/dV$ map using a minimal model Hamiltonian\cite{Koch2006B,Seldenthuis2008} with $\hbar\omega_0$=2.6~meV obtained from the measurements. We consider up to $n=10$ and solve the standard master equation
to find the $dI/dV$ map \cite{Seldenthuis2008}. A good quantitative agreement is found for $\lambda=2.21$. For simplicity, the Hamiltonian includes neither spin degrees of freedom of an electron nor the magnetic moment of the Fe$_4$, which is justified because the parallel lines do not change with an applied magnetic field. The resulting $dI/dV$ map, shown in Figure 2d, reproduces the main features of the measurements, such as the the low-bias gap and the presence of equally spaced lines running parallel to the diamond edges.

The value of $\lambda$ can be independently estimated from the $dI/dV$ peaks ($(dI/dV)_{\rm{max}}$) of the SET excitations as a function of $V$ at a given $V_{g}$\cite{Leturcq2009,Osorio2010}. Figure 3a shows $dI/dV$ versus $V$ at a fixed $V_{g}=-0.711$~V. The background contribution due to direct tunneling between the electrodes is subtracted from the data. The solid line in Figure 3a represents a fit of $(dI/dV)_{max}$ to the FC factor defined in Eq.~(\ref{eq:FC}) with $\lambda=2.05$, which, within the uncertainty, is consistent with the value of $\lambda$ estimated from the low-bias gap in Figure 2. Note that the curve does not fit the low-bias region because Eq.~(\ref{eq:FC}) does not take into account contributions from non-resonant co-tunneling which are important in this region. The average $\lambda$ obtained from different $dI/dV$ traces is $2.0\pm 0.2$ (see Supp.Info.)

\begin{figure}
\includegraphics[width=1\textwidth]{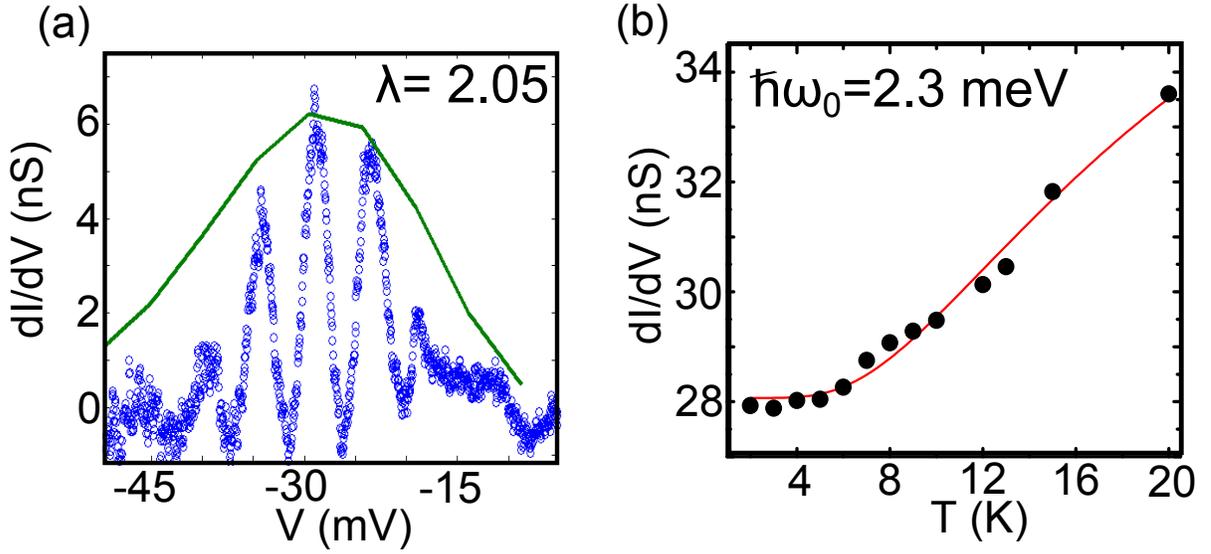}
\caption{\textbf{Analysis of the Franck-Condon blockade.}
(a) Differential conductance measured as a function of $V$ at $V_{g}=-0.711$ V. The background signal due to direct tunneling between the electrodes has been subtracted. The solid line is a fit to $(dI/dV)_{max}$ using the Franck-Condon progression described in the main text. We obtain $\lambda=2.05$, a value that is consistent with the $\lambda$ estimated from the size of the low-bias gap and the value obtained with the rate equations. (b) Temperature dependence of $(dI/dV)_{max}$ measured at $V_{g}=-0.700$ V and $V=12$ mV corresponding to the Coulomb blockade regime. From the fit (solid line) we obtain $\hbar\omega_{0}=2.3$ meV. This energy is consistent with the energy spacing of the excited vibronic states ($\Delta E=2.6$ meV).}
\label{figure3}
\end{figure}

The value of $\omega_{0}$ can also be independently estimated from the temperature dependence of $dI/dV$ in the Coulomb blockade regime. Figure 3b shows $dI/dV$ at different temperatures measured at $V_{g}=-0.700$ V and $V=12$ mV, which corresponds to the Coulomb blockade regime close to the diamond edge at positive bias. Note that $V=12$ mV is greater than the threshold bias to lift the Franck-Condon blockade. The value of $dI/dV$ increases non-linearly with increasing temperature. Such an increase of the conductance with the temperature can be explained with the absorption by the tunneling electrons of one or more vibrational quanta of the molecule\cite{Leturcq2009} (see Figure S5 in the Supp. Info.). Previously forbidden transitions become available and side-bands of sequential tunneling may appear in the Coulomb blockade regime parallel to the Coulomb diamond edges\cite{LeRoy2004,Koch2006,Luffe2008,Leturcq2009}. It is observed in carbon nanotubes\cite{Leturcq2009} that the intensity of the absorption sidebands increases with increasing $T$. If the enhanced tunneling in the Coulomb blockade regime is indeed due to the absorption of vibrons, the temperature dependence obeys Bose-Einstein statistics so that $(dI/dV)_{max} \propto 1/k_{B}T \times 1/(exp(\hbar\omega_{0}/k_{B}T)-1)$\cite{Leturcq2009}. The solid line in Figure 3b is a fit to the experimental data with $\hbar\omega_{0}=2.3$~meV, which is consistent with the energy spacing of the excitations $\Delta E=2.6$ meV. Note that the $dI/dV$ maps measured at 1.8 K do not show evidence of absorption-induced side-bands, indicating a fast vibrational relaxation of the vibrational mode\cite{Luffe2008} in combination with a low $T$ compared with the energy of the vibron ($k_{B}T\ll\hbar\omega_{0}$). Only by increasing temperature we start to observe signatures of peaks within the Coulomb blockade region because vibrons can be excited and their relaxation time becomes longer (see Supp. Info.).

In this work we do not include the effect of oscillations of the center of motion of the Fe$_{4}$ relative to the electrodes, although the frequency could be of the order of meV, as shown for the C$_{60}$ molecule\cite{Park2000}. The oscillations are coupled to the Fe$_{4}$ via displacement-dependent tunneling matrix elements \cite{Koch2004}. We estimate that the coupling strength of the oscillations can not induce Franck-Condon blockade. More details on this issue can be found in the Supp. Info.

So far, the analysis of our experimental data consistently shows that the measured conductance map is due to the FC blockade effect with $\hbar\omega_{0}=2.6$ meV and $\lambda\simeq2.2$. Henceforth, we present results obtained from DFT calculations on the Fe$_{4}$ molecule and compare them with the experimental values. We first find the optimized geometries for the neutral Fe$_{4}$ and singly-charged Fe$_{4}$ molecules using DFT. Then, we calculate the normal modes of the neutral Fe$_{4}$ molecule within the simple harmonic oscillator approximation. In our calculations, we consider only an isolated Fe$_{4}$ molecule without Au electrodes. This is justified because the Fe$_{4}$ molecule is not covalently bonded to the electrodes. For each normal mode, we compute the dimensionless coupling constant\cite{Seldenthuis2008} from
\begin{eqnarray}
\lambda &=& \sqrt{\frac{\omega}{2\hbar}} \Omega^T M (R_0 - R_0^{\prime}),
\label{eq:lambda}
\end{eqnarray}
where $\omega$ and $M$ are the normal-mode frequency and the square diagonal matrix of atomic masses, respectively. $\Omega^T$ is the transpose of the mass-weighted normal-mode column eigenvector with $\Omega^T M \Omega=1$. Here, $R_0$ and $R_0^{\prime}$ are column vectors corresponding to the optimized coordinates of the neutral and charged Fe$_4$ molecules, respectively.

Our DFT calculations show that only three vibrational modes with energies 2.0, 2.5 and 3.7~meV have a value of $\lambda$ greater than unity such as 1.27, 1.33 and 1.46, respectively, as shown in the inset of Figure 4a. The rest of the vibrational modes have a value of $\lambda$ much lower than unity except for several tens of modes with $\lambda$ of the order of 0.1, as shown in Figure 4a. The three vibrational modes have energies close to the experimentally extracted energy values, and the coupling constants are similar to the experimental values. Despite having three normal modes with $\lambda>1$, our transport calculations show only one broad $dI/dV$ peak at the corresponding multiples of the vibrational excitation. This is due to the thermal broadening at 1.8 K (see Fig. S6 in Supp. Info.).
\begin{figure}
\includegraphics[width=0.45\textwidth]{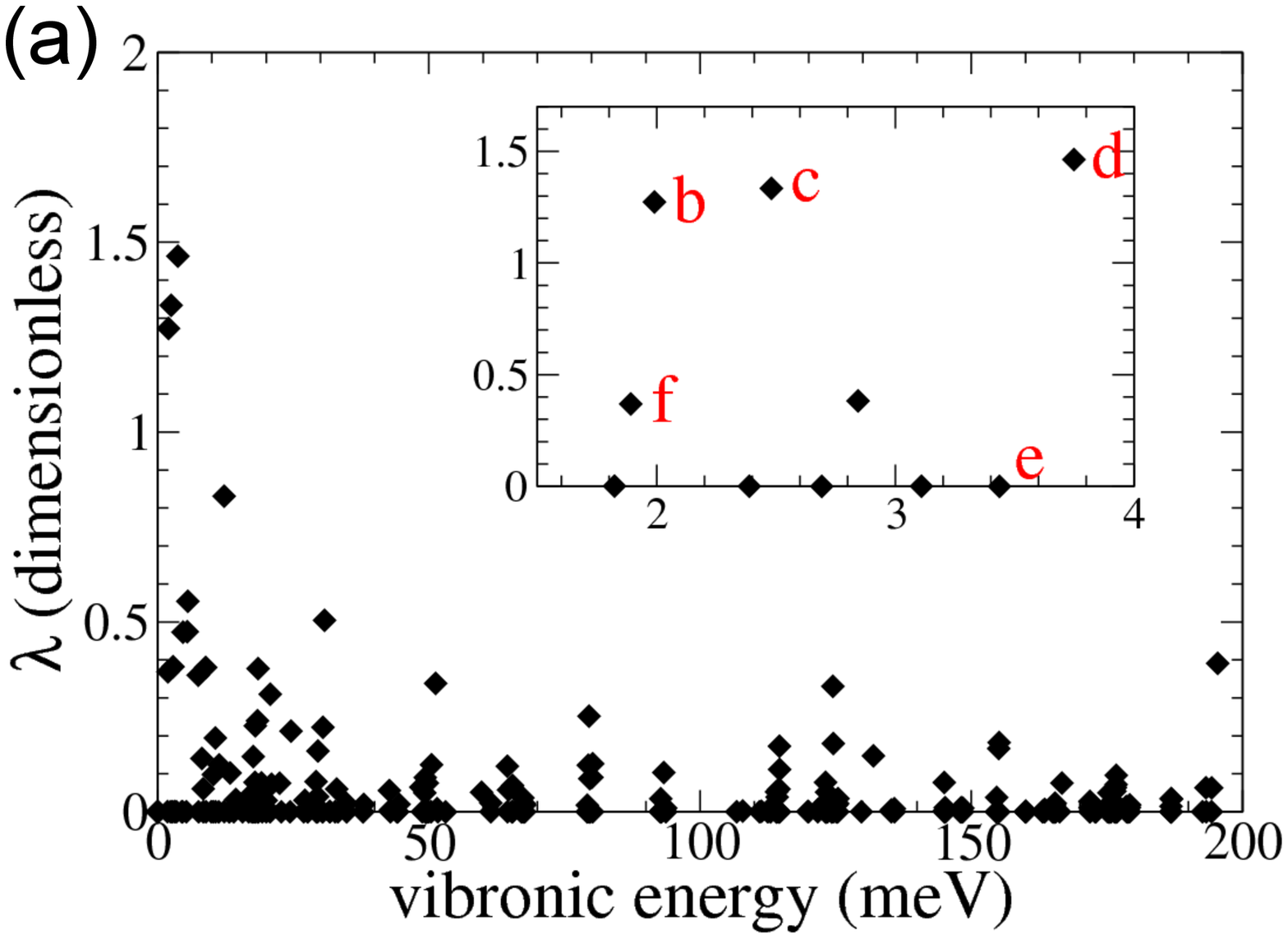}
\includegraphics[width=0.4\textwidth]{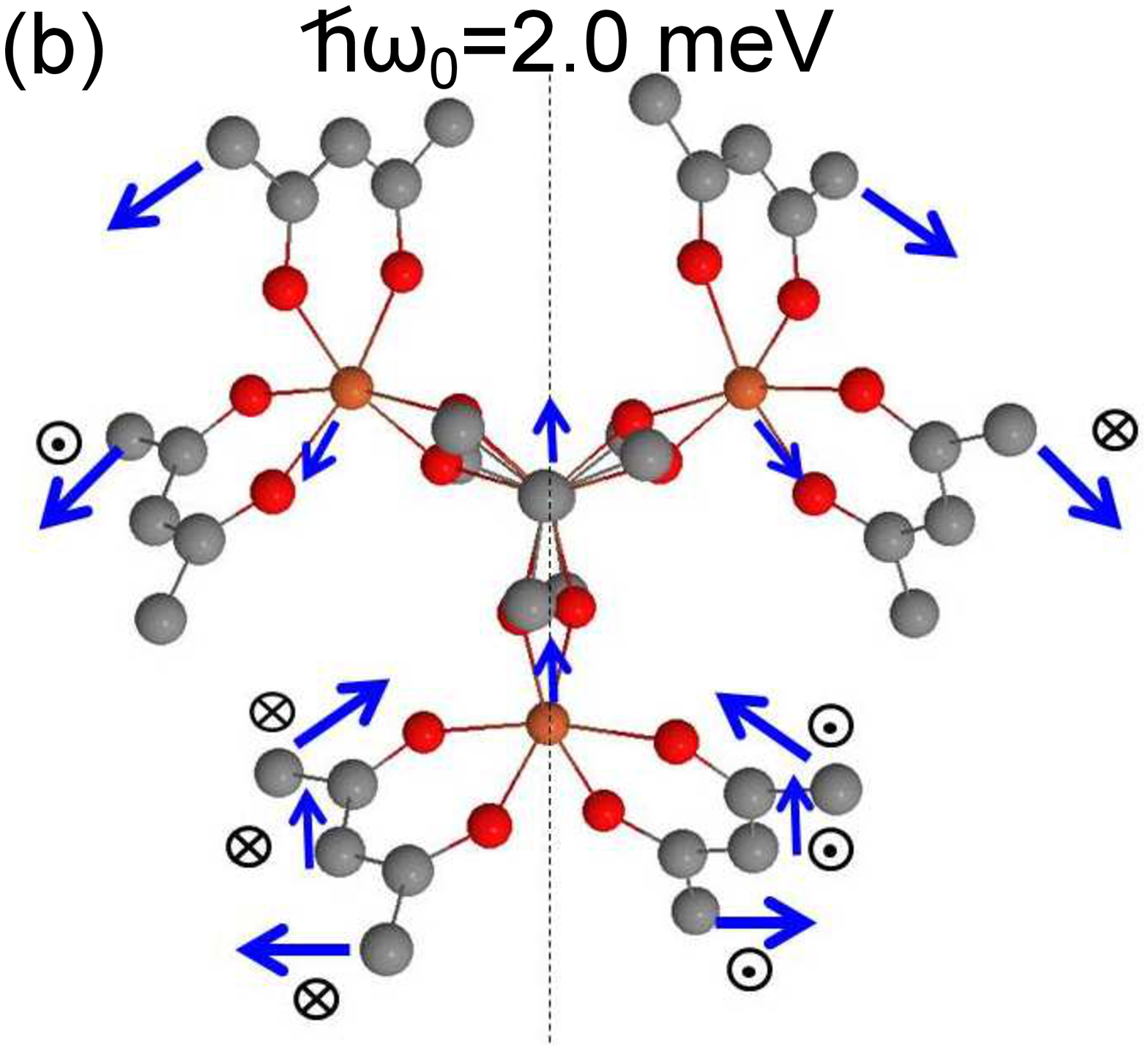}
\includegraphics[width=0.4\textwidth]{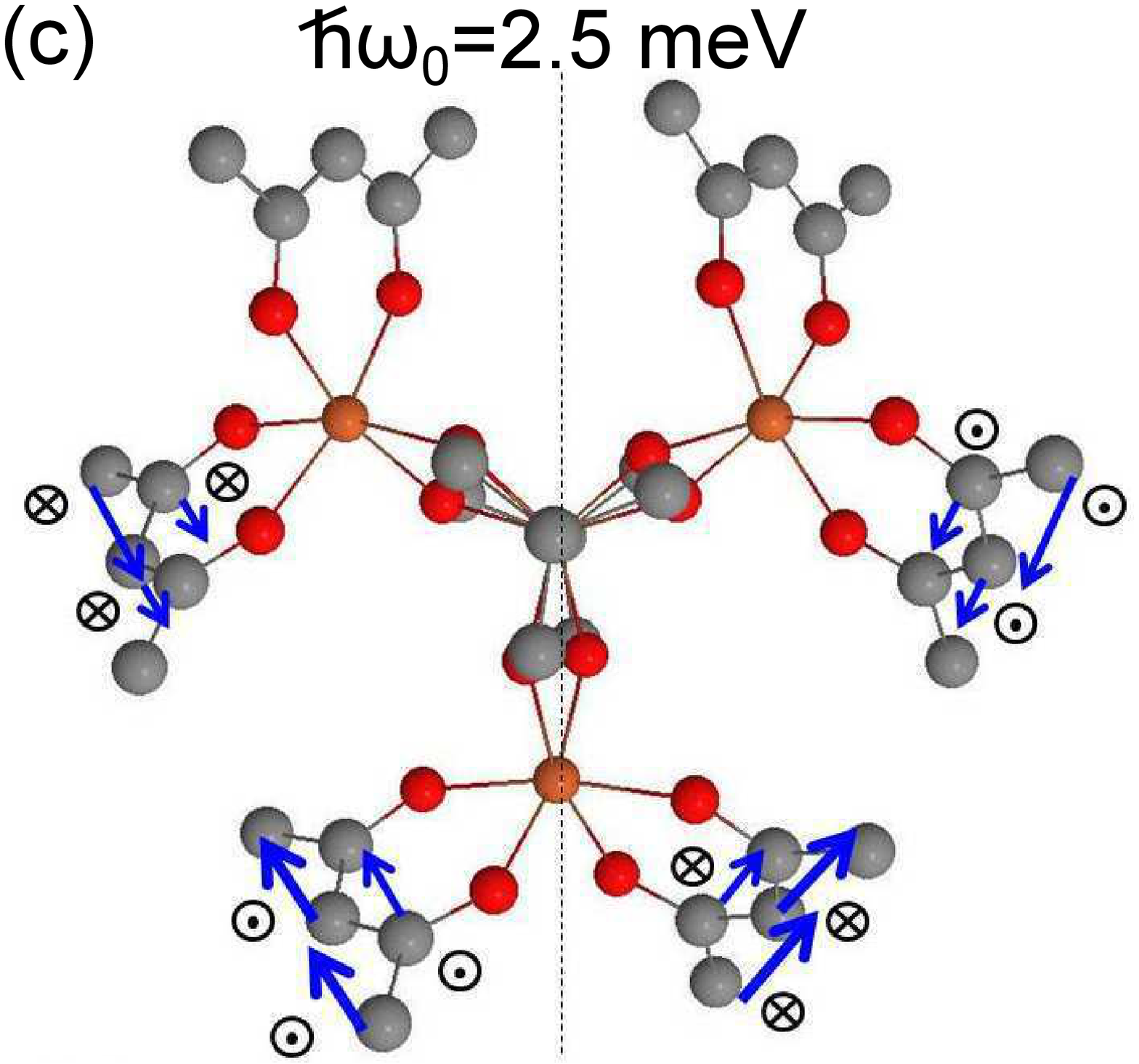}
\includegraphics[width=0.4\textwidth]{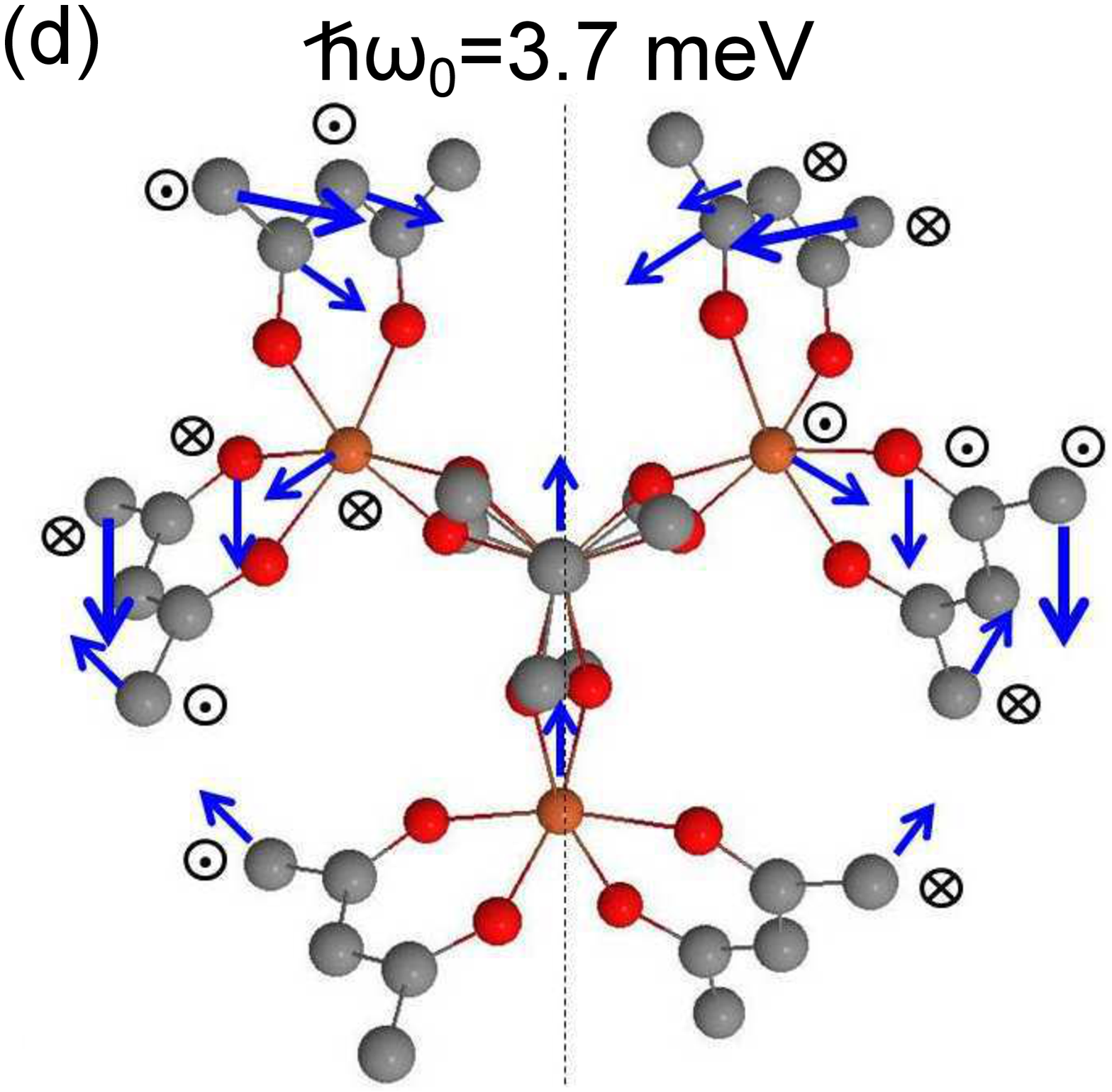}
\includegraphics[width=0.4\textwidth]{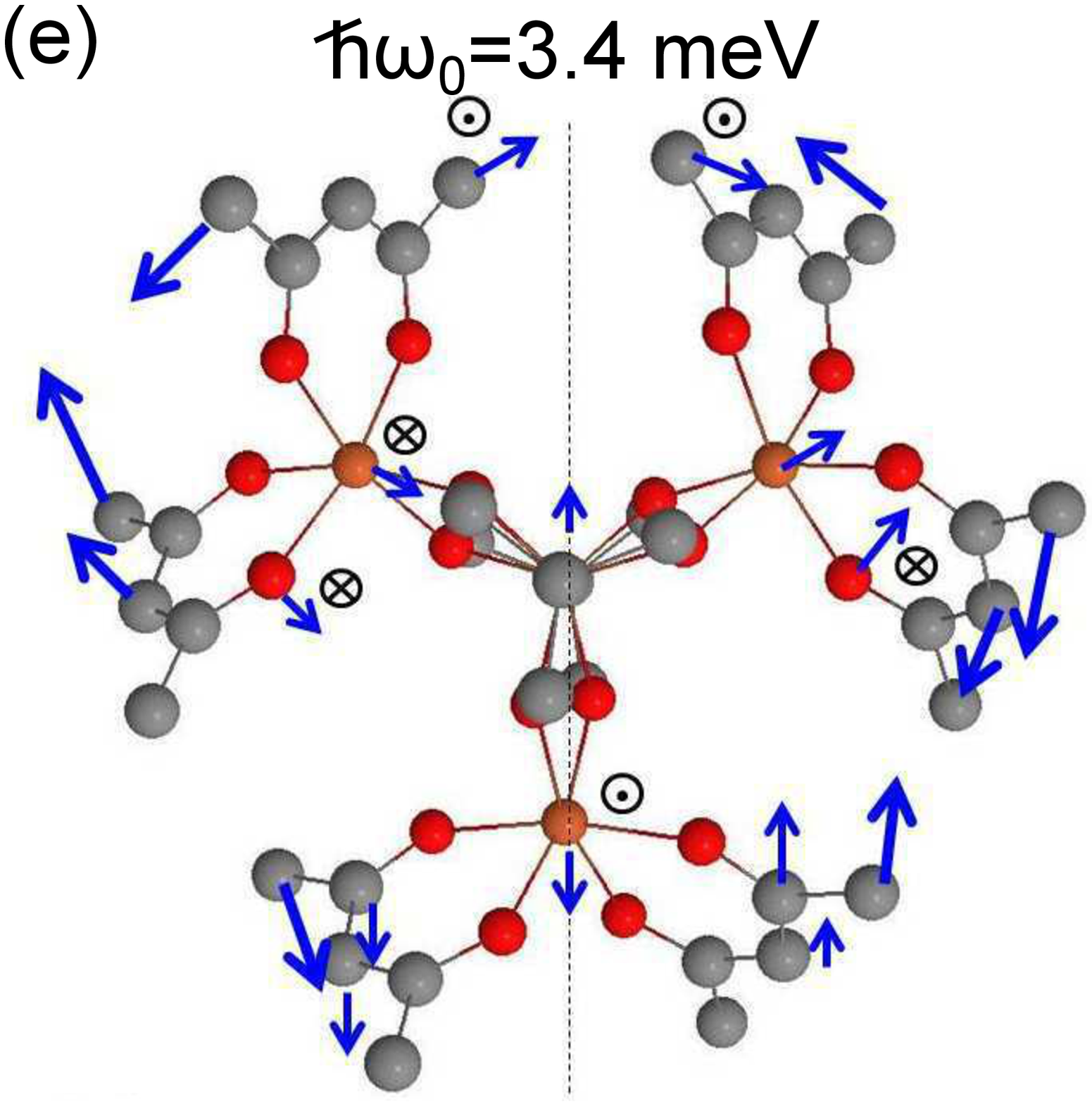}
\includegraphics[width=0.4\textwidth]{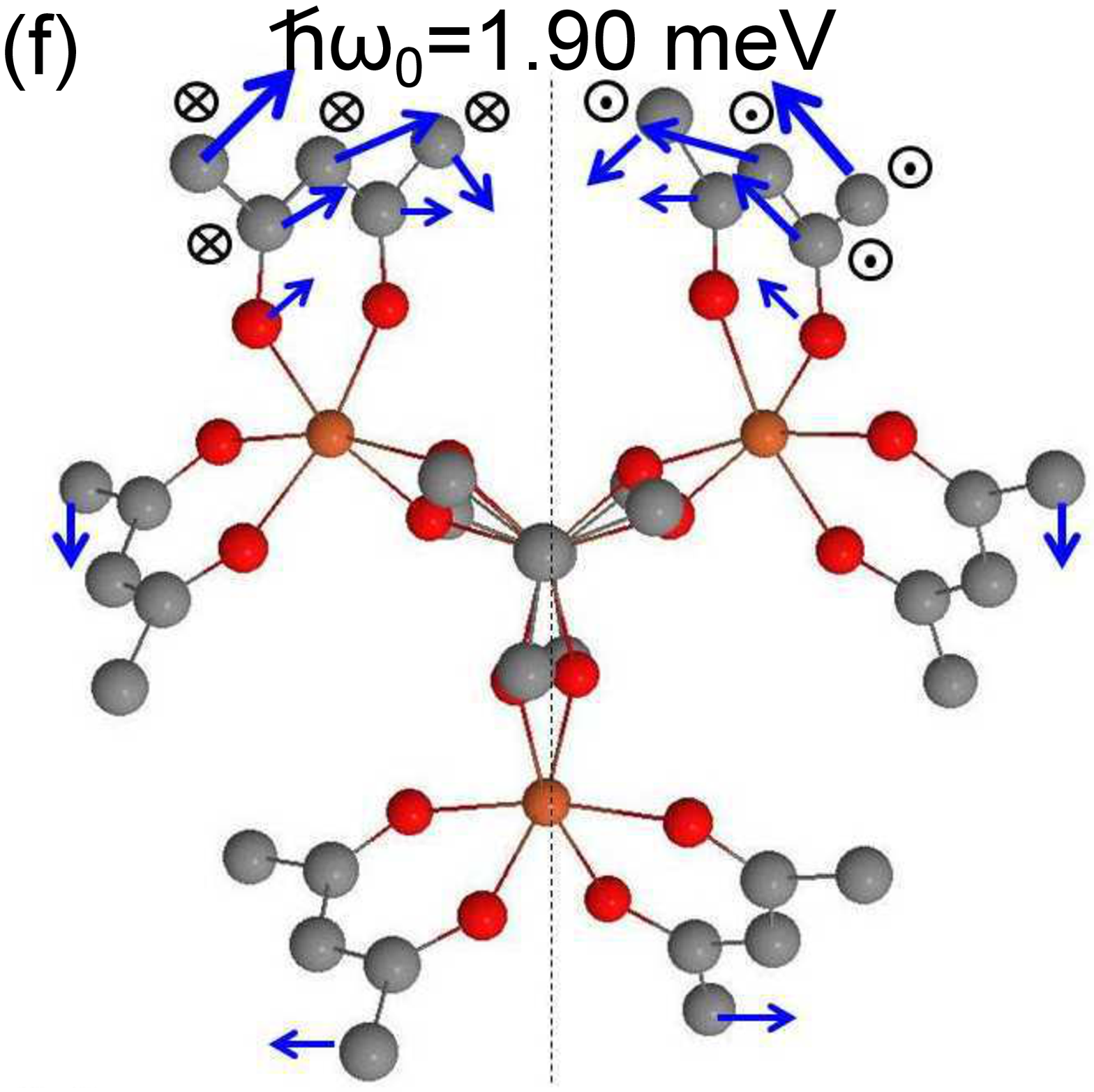}
\caption{\textbf{DFT results for Fe$_{4}$ SMM.}
(a) Calculated electron-vibron coupling constant vs. $\hbar\omega$ for Fe$_{4}$. The inset is a zoom-in of $\lambda$ vs $\hbar\omega$ showing the five normal modes illustrated in (b-f): $\hbar\omega$=2.0, 2.5, 3.7, 3.4, 1.9~meV. The Fe$_4$ molecules in (b-f) are projected onto the $x$-$y$ plane, with the dashed vertical lines indicating the $y$ axis. The color code in (b-f) is Fe (orange), O (red), C (gray). H atoms and phenyl rings are not shown. The length of the arrows represents the magnitude of the displacements. The circled dot and circled cross are displacements along the positive and negative $z$ axes. Only significant displacements are shown.}
\label{figure4}
\end{figure}

We now analyze the characteristics of the vibrational modes in order to understand why only a few of them have a strong coupling constant. The three modes with strong coupling are shown in Figure 4b-d, where the dashed vertical lines represent the $y$ axis. The Fe$_{4}$ molecule of interest has twofold (C$_2$) symmetry about this axis. Movies of the three modes are available in the Supporting Information. According to the group theory\cite{Dresselhaus2008}, all the normal modes of the C$_2$ symmetric Fe$_4$ molecule can be classified into symmetric and antisymmetric modes about the $y$ axis, or  A and B representations, respectively. A half of the non-zero-frequency normal modes belong to the A representation and the other half to the B representation. We find that all the modes in the B representation have a very small value of $\lambda$, which is less than 0.00045 in our numerical accuracy, and that the value of $\lambda$ for the modes in the A representation varies. For example, the normal mode shown in Figure 4e, belongs to the B representation, and it has an electron-vibron coupling constant of 0.0001. There are several distinctive features in the three modes with strong coupling compared to other modes with weaker coupling: (i) The normal modes in Figure 4b-d belong to the A representation, (ii) the major vibrations are from heavier elements such as Fe, O, and C atoms, and (iii) the modes have low frequencies.

These observations concerning the strength of the electron-vibron coupling can be rationalized as follows. The coupling $\lambda$ in Eq.~(\ref{eq:lambda}) is proportional to the inner product between the normal-mode eigenvectors and the difference vector in the equilibrium coordinates of the neutral and charged Fe$_4$ molecules. Considering the twofold symmetry of an isolated Fe$_4$ molecule, it is likely that this symmetry can be preserved even for an Fe$_4$ molecule bridged between electrodes, because the molecule is not chemically bonded to them. If the charged Fe$_4$ molecule has the same twofold symmetry, then the coordinate difference vector also bears the twofold symmetry. This implies that for normal modes in the A representation, each term in Eq.~(\ref{eq:lambda}) contributes to the coupling with the same sign so that the coupling constant can become large. However, when normal modes are in the B representation, terms in Eq.~(\ref{eq:lambda}) cancel out, and the coupling constant becomes very small (i.e. $\lambda \ll 1$). See Figure 4e for an example. Note that this result suggests that a Fe$_4$ molecule chemically bonded to electrodes may not bring strong electron-vibron coupling due to possible broken molecular symmetry. Furthermore, among the normal modes with the twofold symmetry, the coupling is expected to be stronger when the vibrations mainly arise from heavier elements such as Fe, O, and C atoms, rather than H atoms. This is due to the atomic mass term included in Eq.~(\ref{eq:lambda}).

Finally, we discuss possible sources of discrepancy between the experimental data and the DFT calculations. The first source is that the distribution of an extra electron added (or tunneled) to a neutral Fe$_{4}$ molecule has not been experimentally determined. The DFT-calculated value of $\lambda$ depends on the difference in equilibrium coordinates of the neutral and charged Fe$_{4}$ SMMs, as shown in Eq.~(2). Changes in the distribution of the extra electron can significantly change this difference and they may break the molecular twofold symmetry. In our DFT calculations, we consider one particular charge distribution where the extra electron is uniformly distributed over all the four Fe ions. In the case of carbon nanotubes, it was reported that strong electron-vibron coupling is induced when the tunneled electron is localized\cite{Sapmaz2006}. The second source is that the value of $\lambda$ depends on environmental factors\cite{Pasupathy2005,Osorio2007}, yet DFT calculations do not fully capture them, such as image charge effects, localized charge impurities, and various molecular orientations relative to the electrodes. In this direction, note that we do not observe FC blockade in all Fe$_{4}$ devices. Similarly, a variation of $\lambda$ for different devices has also been observed for carbon nanotubes\cite{Osorio2007} and C$_{140}$ molecules\cite{Pasupathy2005}. The aforementioned two sources are correlated and make it difficult to theoretically precisely assign a particular mode as the cause of the FC blockade. A further study along this direction is interesting but beyond the scope of the current work.

In conclusion, we have studied single-electron transport via an individual Fe$_4$ SMM in three-terminal devices. We have observed a suppression of the low-bias conductance and explained its origin with a FC blockade effect caused by strong coupling between the electric charge and a vibrational
mode of the Fe$_4$. From a detailed comparison with the FC model, we extracted the values of the vibron frequency and electron-vibron coupling constant from our experimental data. The values agree with the DFT-calculated results, which also suggest the possibility of increasing the coupling by avoiding chemical bonding at the interface. This is the first direct experimental evidence of the FC blockade effect for a small magnetic molecule with a diameter of about 2~nm. Our findings will stimulate further research on the role of molecular vibrations in molecular transport and more specifically on the impact of such strong electron-vibron coupling on the molecular spin degrees of freedom.


\begin{acknowledgement}
The authors thank J. Seldenthuis, A. Zyazin and J. M. Thyssen for discussions. This work was supported by the EU FP7 program under the Grant Agreement ELFOS, the project 618082 ACMOL and the advanced ERC grant (Mols@Mols). It was also supported by the Dutch funding organization NWO (VENI) and OCW. Y.Y., M.W., X.Z., and K.P. were supported by U.S. National Science Foundation DMR-0804665 and DMR-1206354 SDSC Trestles under DMR060009N and VT-ARC.
\end{acknowledgement}



\end{document}